\documentclass[letterpaper,floatfix,reprint,superscriptaddress,longbibliography,amsmath,amssymb,aps,pra]{revtex4-2}
\usepackage{graphicx,setspace}
\usepackage{braket}
\usepackage{filecontents}

\newcommand{\diff}{\mathop{}\!\mathrm{d}}
\newcommand{\unit}{\mathop{}\!\mathrm}

\begin{document}

\title{Diamond magnetometry and gradiometry towards subpicotesla DC field measurement}%

\author{Chen Zhang}%
\email[]{c.zhang@pi3.uni-stuttgart.de}
\affiliation{3rd Institute of Physics, University of Stuttgart, Pfaffenwaldring 57, Stuttgart 70569, Germany}

\author{Farida Shagieva}%
\affiliation{TTI GmbH / TGZ SQUTEC, Nobelstraße 15, Stuttgart 70569, Germany}

\author{Matthias Widmann}%
\affiliation{3rd Institute of Physics, University of Stuttgart, Pfaffenwaldring 57, Stuttgart 70569, Germany}

\author{Michael Kübler}%
\affiliation{3rd Institute of Physics, University of Stuttgart, Pfaffenwaldring 57, Stuttgart 70569, Germany}

\author{Vadim Vorobyov}%
\affiliation{3rd Institute of Physics, University of Stuttgart, Pfaffenwaldring 57, Stuttgart 70569, Germany}

\author{Polina Kapitanova}%
\affiliation{Department of Physics and Engineering, ITMO University, Saint Petersburg 197101, Russia}

\author{Elizaveta Nenasheva}%
\affiliation{Giricond Research Institute, Ceramics Co. Ltd., Saint Petersburg 194223, Russia}

\author{Ruth Corkill}%
\affiliation{Institute for Modelling and Simulation of Biomechanical Systems, Paffenwaldring 5a, Stuttgart 70569, Germany}

\author{Oliver Röhrle}%
\affiliation{Institute for Modelling and Simulation of Biomechanical Systems, Paffenwaldring 5a, Stuttgart 70569, Germany}

\author{Kazuo Nakamura}%
\affiliation{Leading-Edge Energy System Research Institute, Fundamental Technology Dept., Tokyo Gas Co., Ltd., Yokohama 230-0045, Japan}

\author{Hitoshi Sumiya}%
\affiliation{Advanced Materials Labotratory, Sumitomo Electric Industries, Ltd., Itami 664-0016, Japan}

\author{Shinobu Onoda}%
\affiliation{Takasaki Advanced Radiation Research Institute, National Institutes for Quantum and Radiological Science and Technology, Takasaki 370-1292, Japan}

\author{Junichi Isoya}%
\affiliation{Faculty of Pure and Applied Sciences, University of Tsukuba, Tsukuba 305-8573, Japan}

\author{Jörg Wrachtrup}%
\email[]{j.wrachtrup@pi3.uni-stuttgart.de}
\affiliation{3rd Institute of Physics, University of Stuttgart, Pfaffenwaldring 57, Stuttgart 70569, Germany}


\begin{abstract}
	Nitrogen vacancy (NV) centers in diamond have developed into a powerful solid-state platform for compact quantum sensors. However, high sensitivity measurements usually come with additional constraints on the pumping intensity of the laser and the pulse control applied. Here, we demonstrate high sensitivity NV ensemble based magnetic field measurements with low-intensity optical excitation. DC magnetometry methods like, e.g., continuous-wave optically detected magnetic resonance and continuously excited Ramsey measurements combined with lock-in detection, are compared to get an optimization. Gradiometry is also investigated as a step towards unshielded measurements of unknown gradients. The magnetometer demonstrates a minimum detectable field of $0.3 - 0.7$ pT in a 73 s measurement by further applying a flux guide with a sensing dimension of 2 mm, corresponding to a magnetic field sensitivity of $2.6 - 6 \unit{pT/\sqrt{Hz}}$. Combined with our previous efforts on the diamond AC magnetometry, the diamond magnetometer is promising to perform wide bandwidth magnetometry with picotesla sensitivity and a cubic-millimeter sensing volume under ambient conditions.
\end{abstract}

\maketitle
\section{Introduction}

Quantum sensors have made extraordinary progress in sensitivity, precision, bandwidth, spatial and temporal resolution over the past years \cite{RN1,RN2,RN3,RN4}. It has enabled measurements with cutting-edge performance for various physical quantities, including frequency standard, magnetic and electric field, temperature, rotation, and gravitational field \cite{RN5,RN6,RN7,RN8,RN9,RN10}. Among all the efforts dedicated to advancing measurement limits, measurements of static or low-frequency magnetic fields (DC magnetometry) belong to the most important ones \cite{RN11,RN12}. DC magnetometry allows a wide range of applications in fields such as medical and material science \cite{RN13,RN14,RN15,RN16,RN17,RN18}.

Highly-sensitive DC magnetometry techniques have been developed across various platforms, with sensitivities from femto- to nanotesla \cite{RN18,RN19,RN20,RN21, RN39}. Superconducting quantum interference devices (SQUID) were holding the sensitivity record of $1 \unit{fT/\sqrt{Hz}}$ with spatial resolution of a few centimeters \cite{RN22}, until atomic vapor cells achieved a record sensitivity of $160 \unit{aT/\sqrt{Hz}}$ in a volume of 0.45 cm$^3$ with gradiometry measurement \cite{RN3,RN23}. However, there are still technical challenges concerning further miniaturization without compromising such high sensitivities \cite{RN24}. On the other hand, the nitrogen-vacancy center in diamond has evolved into a competitive room-temperature platform for magnetometry with its exceptional spatial resolution, dynamic range, and sensitivity \cite{RN2,RN25}. With neither cryogenic nor heating requirements in operation, the sensor head can be reduced to the size of the diamond itself, i.e. sub-mm$^3$, still keeping the pT sensitivity \cite{RN2}. Furthermore, the NV$^-$ centers can be operated in bias fields from zero to a few Tesla \cite{RN26,RN27}. In 2016, a $15 \unit{pT/\sqrt{Hz}}$ DC sensitivity was demonstrated, and the magnetometer was successfully applied in sensing the neuron activity from a marine worm \cite{RN28}. Such a high-sensitive NV magnetometry usually requires high excitation laser power. Recently, by applying a flux concentrator (FC) to enhance the local magnetic field by more than a factor of 200, the DC sensitivity has been further improved from $300 \unit{pT/\sqrt{Hz}}$ to $0.9 \unit{pT/\sqrt{Hz}}$ while the excitation laser power is significantly lower than the power commonly used in high sensitivity measurements \cite{RN29}. Although this improvement in sensitivity comes with a price of reduced spatial resolution, it still has a $\unit{cm}^3$ sensing volume at a significantly higher dynamic range compared to optically pumped magnetometers (OPM).

In this work, we present DC magnetometry and gradiometry with (0.5 mm)$^3$ diamond volume, which contains NV center ensembles with a minimum ODMR linewidth of 28 kHz and a long dephasing time of 8.5 $\unit{\mu s}$. With excitation laser power below 100 mW, multiple diamond DC magnetometry methods are investigated. Pulsed schemes with continuous excitation and readout are demonstrated with a better sensitivity potential than optimized CW-ODMR measurements. On the other hand, we also show that the CW-ODMR driving both hyperfine and double resonance (DR) transitions demonstrate optimized sensitivity close to Ramsey measurements with the same low-power laser excitation. We present a measurement showing the minimum detectable field at $2 - 3$ pT within the bandwidth of $0 - 200$ Hz in 73 s measurement time, and the 1 Hz normalized noise spectral density is 17 $\unit{pT/\sqrt{Hz}}$. We also present a gradiometry measurement, showing the minimum detectable differential field noise at $4 - 6$ pT with the 73 s measurement time. Furthermore, by applying a ferrite flux guide (FG), the measured minimum detectable field goes down to subpicotesla at $0.3 - 0.7 \unit{pT}$ over the mentioned bandwidth, corresponding to a sensitivity level at $2.6 - 6 \unit {pT/\sqrt{Hz}}$.

\section{\label{sec II}Experiment}
\subsection{Experimental setup and methods}
The optimization of diamond magnetometry sensitivity has been reviewed in detail recently \cite{RN30}. NV$^-$ magnetometry is based on the Zeeman shift of its ground state sublevels. The S = 1 spin can be initialized by green laser pumping, and the triplet ground state can be resonantly driven by microwaves (MW). The typical way of measuring the Zeeman shift is to detect the fluorescence change induced by the sublevel population difference. The intrinsic noise of the magnetometer can be defined as $\delta B=\sigma (t)/(\diff S / \diff B)$, where $\sigma(t)$ is the detected noise floor excluding the external field noise, and the scalar factor $\diff S/\diff B$ is the change in signal d$S$ occurring per magnetic field change d$B$. In NV$^-$ ensemble magnetometry, the noise characteristic is essentially determined by the characteristics of the NV$^-$ ensembles. The minimum detectable field scales with the sensing volume as well as the measurement time \cite{RN4}. In this work, by applying an optimized diamond sample with narrow linewidth NV$^-$ ensembles, we mainly focus on compromising the signal level and the technical noise and feasibility.

\begin{figure}
	\includegraphics{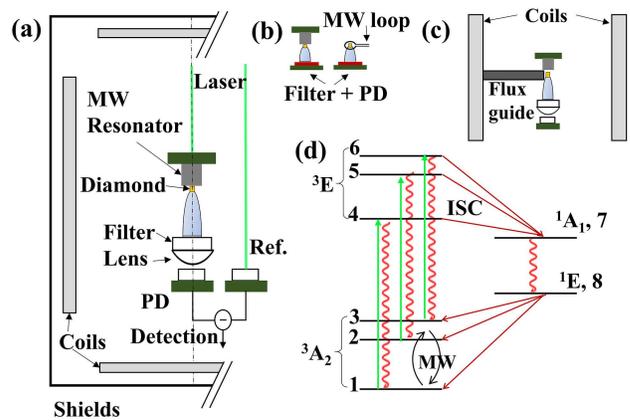}
	\caption{\label{fig1}(a) The experimental setup is installed in a magnetic field shield for high sensitivity measurements. (b) Two identical diamond sensors are constructed to measure the noise floor in the gradiometer configuration. The reference channel uses a loop antenna. (c) A flux guide is placed between the diamond and a coil for enhancement of the magnetic field sensitivity. (d) NV$^-$ center energy diagram.}
\end{figure}

Figure \ref{fig1} shows the experimental setup and the energy level diagram for the photo- and spin kinetic simulations \cite{Note}. A magnetic shield cube entirely encloses the setup to attenuate the magnetic field noise in the lab. The optical compound parabolic concentrator (CPC) is used for high fluorescence collection efficiency, which exceeds $60\%$ \cite{RN4}. The dielectric resonator antenna generates a uniform driving MW field for the NV ensemble \cite{RN31}. We investigate the performance of a gradiometer by constructing two identical diamond sensors, as shown in Fig. \ref{fig1}(b). A loop antenna is applied in the reference channel. In order to test the sensitivity enhancement by the FG, a ferrite rod (MN60) is placed between the diamond and a coil applying the test field, as shown in Fig. \ref{fig1}(c). The diameter of the FG tip is 2 mm. The gap between the FG and the diamond is roughly 1 mm, limited by the CPC lens structure. The FG-assisted enhancement could be higher with its decreasing distance to the diamond.

\begin{figure*}[t]
	\includegraphics[scale=1]{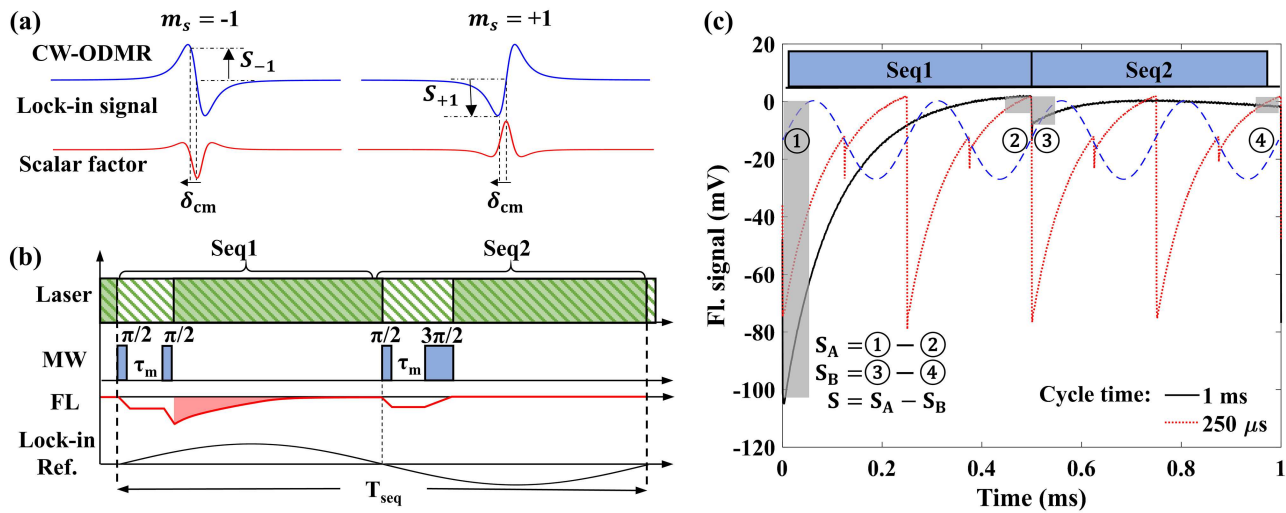}
	\caption{\label{fig2}
		(a) Lock in detected CW-ODMR signal profile and the corresponding scalar factor. $\delta_{cm}$ is a common-mode line shift of the two resonant lines. $S_{-1}$ and $S_{+1}$ are the corresponding signals with the same amplitude but different signs. (b) The CE-Ramsey sequence is used as a promotion of the CW-ODMR measurement. The solid laser blocks depict the laser pulses in a traditional Ramsey measurement, while in CE-Ramsey the laser is continuously applied in all the time bins as the slashes show. The FL curve (red line) depicts the fluorescence signal when the external field is zero. The signal in the time bin corresponding to the solid laser block is the valid signal for magnetic field sensing, as shown in the shadow area in Seq 1. The rest signal becomes negligible when $\tau_m \ll T_{seq}$, which can be seen in (c), fluorescence readout with cycle time $T_{seq} = 1 \unit{ms}$ (black solid curve) and 250 $\unit{\mu s}$ (red dotted curve). The sketched sequence blocks inset correspond to the 1 ms cycle signal. The figure shows both the conventional readout method (shown with the 1 ms cycle signal) and the lock-in readout method (shown with the 250 $\unit{\mu s}$ cycle signal). Signals in the gray areas are calculated following the equations shown in the figure for conventional readout. In lock-in detection, the demodulation reference is applied according to the dashed sinusoidal line.}
\end{figure*}

One of the most critical techniques in sensitive measurements is lock-in detection to avoid low-frequency $1/f$ noise. Figure 2(a) shows the typical CW-ODMR output from the lock-in amplifier (LIA), with the NV centers spin transitions driven by a modulated MW field. The MW modulation phases are of opposite sign for the $\ket{0}\rightarrow\ket{+1}$ and $\ket{0}\rightarrow\ket{-1}$ transitions \cite{RN29}. Thus, the signals $S_{-1}$ and $S_{+1}$, which are induced by the common-mode line shift $\delta _{cm}$, have the same amplitude but different signs. The common-mode line shifts are induced mainly by thermal fluctuations. By tracking the resonance lines in cases of larger magnetic fields, one can maintain the maximum scalar factor and the linearity of the system.

Figure \ref{fig2}(b) shows a Ramsey sequence as an example for pulsed measurement sequences using a LIA, while Fig. \ref{fig2}(c) depicts the signal dynamics together with the acquisition methods. In the up to now used Ramsey measurements \cite{RN2}, the laser is turned off during the field acquisition time $\tau_m$. The fluorescence signal is therefore gated and appears as pulses. However, the signal $\diff{S}$ only amounts to a few percent of the overall signal. The background fluorescence consumes most of the bits of the digitizer when using normal photodetectors. An alternative could be to use a balancing detector. However, when the two channels are not perfectly balanced due to stability reasons, the left pulsed signals still induce overshoot and delay errors in the electronics. Besides, the thermal dynamics induced by the laser pulses contribute to additional errors in the measurement. Thus, we have modified the sequence to a continuous excitation mode with the laser turned on all the time, as shown in Fig. \ref{fig2}(b). The fluorescence during $\tau_m$ and MW pulses time can be neglected due to the low repolarization rate under the low-power laser excitation. One of the apparent advantages of using continuous excitation is the simplicity of the laser optics, e.g., an optical switch is no longer required even for MW pulsed measurements. By simply applying AC coupling to the digitizer, any background signal can be removed without noises induced by the pulsed signals. Moreover, the photodetectors do not need wide bandwidth under continuous excitation for the high dynamic pulse signals so that the gain could increase for a better signal-noise ratio (SNR). Heating fluctuation induced by the pulsed laser can be removed, and thermal stability can be improved by monitoring the laser power drifts for compensation. The disadvantage is that the laser repolarizes the NV center during $\tau_m$, which reduces $\diff{S}$. In a quantitative treatment of the measurements, this is treated as a lower polarization rate, i.e., it reduces the contrast a bit but does not induce error to the phase accumulation. We would like to note that the continuous excitation mode can also be applied in other pulsed measurement schemes, such as Hahn-echo and high order dynamical decoupling sequences. The black curve in Fig. \ref{fig2}(c) is the fluorescence signal for a cycle time of 1 ms. The labeled grey areas and the inset formulas describe the acquisition method which used in previous work \cite{RN4}. For the sensitivity calculation using the signal S, we need to consider the scaling of noise with the factor $\sqrt{T_{seq}/\Delta t}$, where $T_{seq}$ is the sequence time, and $\Delta t$ is the total acquisition time. On the other hand, the lock-in detection is equivalent to a method, in which $S_A$ and $S_B$ are the averaged signal level of each half reference cycle, and $S$ is the amplitude of the entire cycle. The LIA integrates all the detected fluorescence so that there is no acquisition time intermittency, which deteriorates the sensitivity. Therefore, the measured minimum detectable field, i.e., shot noise of the Ramsey measurement can be expressed as \cite{RN2,RN30}
\begin{equation}
	\delta B_{Ramsey} = \frac{\hbar}{g\mu_B} \frac{1}{C_{det}\sqrt{N\tau_m t}} \sqrt{\frac{T_{seq}}{\tau_m}}
	\label{eq1},
\end{equation}
where $\tau_m$ is optimized from the dephasing time $T_2^*$, $C_{det}$ is the detected contrast, $N$ is the number of photons collected per measurement, and $t$ is the measurement time. Experimentally $N$ can be estimated by the photon detection rate $R = N / T_{seq}$. The sensitivity is calculated by $\eta=\delta B\sqrt t$. The equation calculates the shot noise without explicitly including the photon detection efficiency. When the laser power is not saturating the NV ensembles, as in our case of low optical pumping, the collected photon rate is also related to the pumping power. In Fig. \ref{fig2}(c) the dotted line shows the fluorescence signal with 250 $\unit{\mu s}$ cycle time, and the dashed line is the corresponding lock-in reference. When the time duration for each sequence part is significantly shorter than the repolarization time of the spins, the contrast reduces dramatically due to insufficient repolarization.
The contrast increases with higher pumping power when polarization time $T_p$ is fixed. In this case, $T_{seq}$ can be shorter to make the magnetometer more sensitive. However, latterly we will see that laser power of Watts is required to get a significant improvement of sensitivity. In order to make it more clear in the model, we denote the contrast as $C(P, T_p)$, where $P$ is the applied laser power.
Note that the contrast also decays with the phase factor $\exp{(-\tau_m/T_2^*)}$, the detected contrast $C_{det}=C(P,T_p)\exp{(-\tau_m/T_2^*)}$. 
Meanwhile, improving the laser power leads to higher fluorescence signals due to more optical cycles, and it also contributes to a linear rising in laser noise. It becomes rather complicated to get better sensitivity in practical then. 
Thus, we optimize with a fixed low laser power for the trade-off between the technical noise floor and the signal contrast, which both decrease with higher demodulation frequencies.

\subsection{Parameters optimization}
\begin{figure*}[htb]
	\includegraphics[scale=0.95]{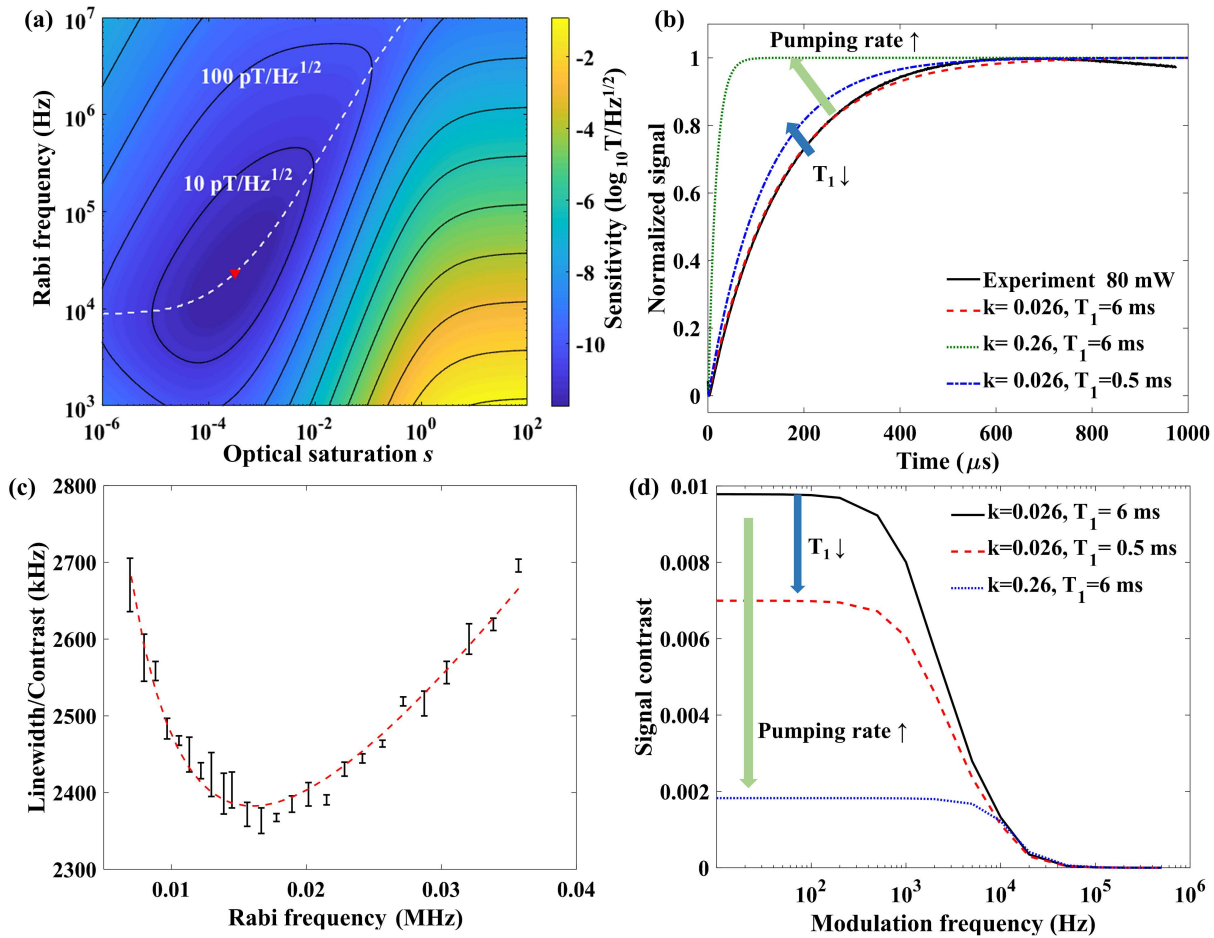}
	\caption{\label{fig3} 
		(a) Theoretical shot noise limited sensitivity optimization with laser pumping and MW driving strength based on the CW-ODMR model. $T_2^* = 8.5 \unit{\mu s} $ and $T_1 = 6 \unit{ms}$ are the parameters used in the calculation. The dashed line shows the optimized MW strength and sensitivity for different excitation powers. The triangle marks the best theoretical sensitivity. 
		(b) The repolarization curve of both the experimental result and the simulation, indicating the effects of the two primary parameters: pumping rate ($k$ in MHz) and $T_1$. 
		(c) The experimental sensitivity parameter linewidth/contrast with different Rabi frequencies. The Rabi frequency for the CW-ODMR method is optimized as 17 kHz. 
		(d) Simulated CW-ODMR signal contrast upon different modulation frequencies (in logarithm). The pumping rate and T1 also affect the bandwidth and contrast of the magnetometer. The diamond magnetometer has a 3 dB bandwidth at around 1.5 kHz, and the signal contrast reduces from 1 kHz to 20 kHz.}
\end{figure*}

Among all the diamond DC magnetometry schemes, Ramsey magnetometry is among the most sensitive methods. However, its practical sensitivity suffers from technical limitations. For example, enormous laser power is needed for the initialization of the spin ensembles. It brings heating problems and extra noise with it. In contrast, CW-ODMR is a method that can be easily implemented with low laser power. In the following, we first analyze the CW-ODMR sensitivity and the optimal laser and MW parameters. Then, Ramsey magnetometry is investigated based on the same laser power. The shot noise measured by the CW-ODMR method is described as \cite{RN32}
\begin{equation}
	\delta B_{CW} = P_F\frac{h}{g\mu_B} \frac{\nu(s,\Omega_R)}{C_{det}(s,\Omega_R)\sqrt{Rt}}
	\label{eq2}.
\end{equation}
Here $P_F$ is a lineshape dependent factor, which is 0.77 for the Lorentzian line profile in this work. The critical parameters for experimental optimization are the linewidth $\nu$ and the contrast $C_{det}$, which are dependent on the Rabi frequency $\Omega_R$, and the ratio of the applied laser power to the saturation power, $s$. Figure \ref{fig3}(a) plots the sensitivity dependence on the parameters. $T_1$ and $T_2^*$ used in the calculation are 6 ms and 8.5 $\unit{\mu s}$, respectively. The optimized magnetometry parameters are $\Omega_R = 23 \unit{kHz}$, and $s = 3\times10^{-4}$. The shot noise limited sensitivity is 2.86 $\unit{pT/\sqrt{Hz}}$ for the signal of a single hyperfine line. By driving the three hyperfine lines, the contrast ideally increases three times. However, the contrast enhancement factor in the experiment is smaller because of reasons such as the near-resonant driving induced by the multi-MW-frequencies. The effect could be limited by applying weak MW driving. In our experiments, the contrast enhancement factor is 2.67. The scalar factor is further improved by DR driving, and the experimental enhancement factor is 1.3 times. Therefore, the sensitivity limit of the CW-ODMR method upon driving all the hyperfine lines and using DR driving is expected to be 0.82 $\unit{pT/\sqrt{Hz}}$. The estimated sensitivity limit based on calculations in Fig.\ref{fig3}(a) does not include the influence of MW modulation. MW modulation reduces the contrast due to the magnetometer frequency response, depending on the optical pumping. In the following, we will discuss the magnetometry parameters at the near optimized optical pumping power. 

In order to have an optimized SNR, the collected fluorescence emission power should be on the order of mW with roughly $3\times10^{-4}$ lower than the saturation laser power. In order to avoid applying an unnecessarily high laser power, the excitation power is optimized by comparing the experimental pumping rate with the simulation of the excitation kinetics. With a fluorescence emission rate of 66 MHz \cite{RN33,RN34}, the optimized pumping rate is estimated to be around 0.02 MHz. This parameter underestimates the pumping rate due to neglecting charge state conversion in the kinetics \cite{RN35,RN41}. We apply a near-optimal laser power with the pumping rate $k = 0.026 \unit{MHz}$. Figure \ref{fig3}(b) plots the experimental fluorescence recovery curve and the simulated results with different parameters. The low pumping rate is the main reason for the millisecond repolarization time. Hence, in the low pumping rate case, the influence from $T_1$ appears on the fluorescence recovery. We take the parameter linewidth/contrast to optimize the MW power, as plotted in Fig. \ref{fig3}(c). We find the optimal $\Omega_R$ to be around 17 kHz, while the applied laser power is roughly 80 mW with the beam size same as the diamond dimension of 0.5 mm. In order to investigate the contrast response to the MW modulation frequency, i.e., the magnetometer bandwidth, the calculated fluorescence responses to different MW modulation frequencies are plotted in Fig. \ref{fig3}(d). The demodulated output magnitude of the fluorescence drops from 1 kHz to 20 kHz modulation frequency by roughly 10 times due to the optical pumping dynamics and $T_1$. Thus, with the MW modulation frequency higher than 1 kHz, the optimal sensitivity also degrades accordingly. The 3 dB bandwidth of the magnetometer is around 1.5 kHz at a pumping rate of 0.026 MHz. It increases to 7.4 kHz with a lower contrast when the pumping rate is 10 times higher. In addition, with a roughly 10 times shorter $T_1$, the contrast also decreases. However, the 3 dB bandwidth only increases to 1.7 kHz. The SNR can be improved by MW modulation within the 3 dB bandwidth since the technical noise floor decreases with the increasing demodulation frequency. After the 3 dB frequency point, it is worth investigating if the SNR can be further improved by reducing both the signal amplitude and the noise, which is discussed below.

\section{\label{sec III}Results and discussions}

\begin{figure*}[hbt!]
	\includegraphics[scale=0.95]{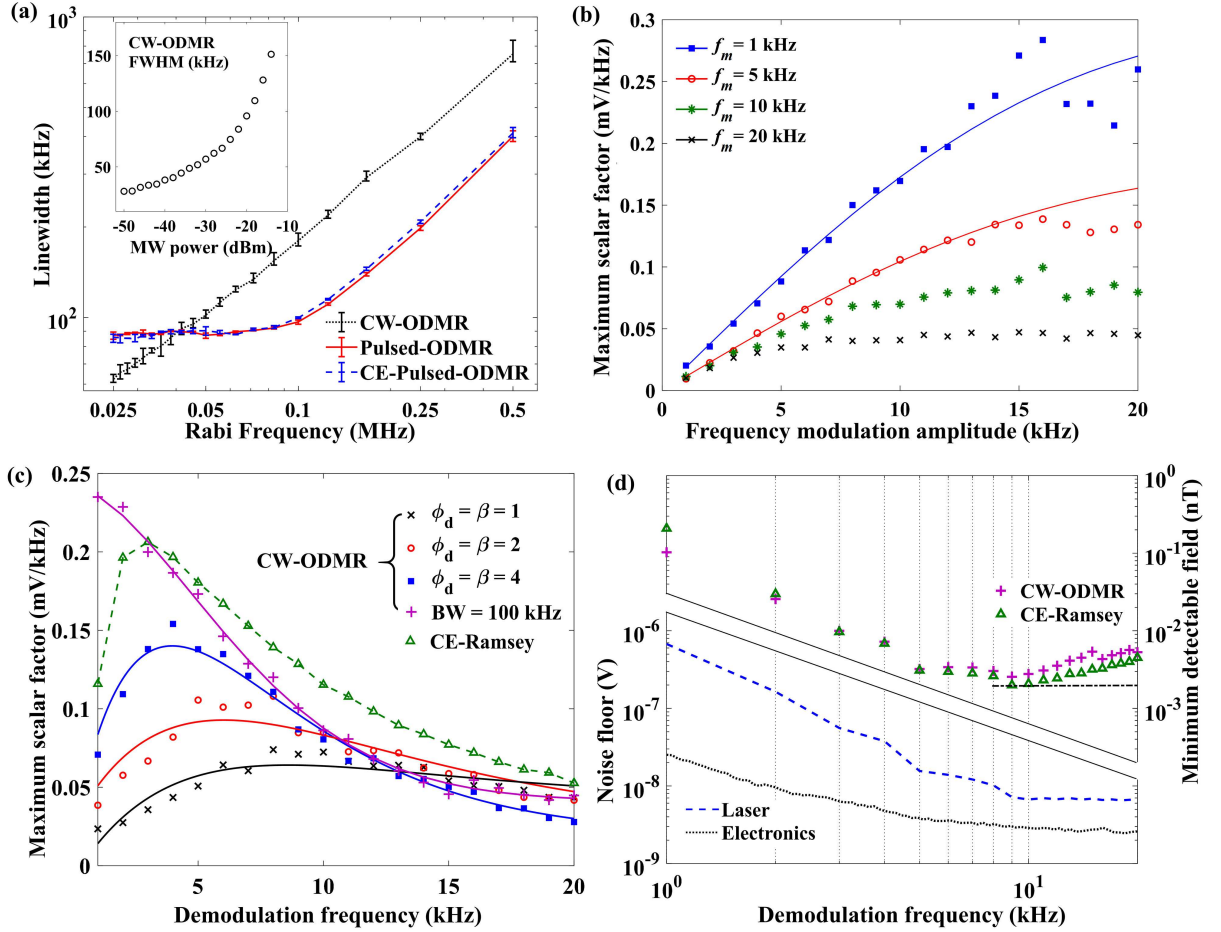}
	\caption{\label{fig4} 
		(a) The linewidth measured with different Rabi frequencies by CW-ODMR (black dotted line), pulsed ODMR (red solid), and CE-pulsed-ODMR (blue dashed line). Both axes are logarithmic. The inset is the CW-ODMR linewidth for decreasing MW power. 
		(b) The scalar factor of CW-ODMR measurement with FM signal $f_m$ = 1 kHz, 5 kHz, 10 kHz, and 20 kHz as a function of the modulation amplitude $f_d$. The data of $f_m$ = 1 kHz and 5 kHz fit well with the derivative of Lorentzian lineshape. However, the fitting cannot work (higher than the measured results) when the modulated MW signal bandwidth $2(f_d + f_m)$ approaches the ODMR linewidth. 
		(c) Comparison of the scalar factor of CW-ODMR (MW phase modulated) and CE-Ramsey with different modulation frequencies. The demodulation frequency of CE-Ramsey measurement is $1/T_{seq}$. The CW-ODMR traces are plotted with different $\phi_d$. For comparison, an extra CW-ODMR curve is plotted with the FM MW signal bandwidth fixed at 100 kHz. 
		(d) plots the noise floor (root mean square noise in V, lower left $y$-axis) of the electronics and laser intensity demodulated from 1 kHz to 20 kHz. Both axes are logarithmic. The frequency resolution of the spectra is 13.7 mHz regarding the 73 s measurement time. In the upper part, the minimum detectable fields (right $y$-axis) of CE-Ramsey and CW-ODMR with $BW$ = 100 kHz are plotted, respectively.}
\end{figure*}

Figure \ref{fig4}(a) shows a comparison of the linewidth between CW-ODMR, pulsed-ODMR, and continuously excited pulsed-ODMR (CE-pulsed-ODMR) at different Rabi frequencies. The diamond demonstrates a narrow CW-ODMR linewidth of 28 kHz FWHM with a weak MW driving, while the calculated minimum linewidth of the diamond with $T_2^*=8.5 \unit{\mu s}$ is $<20$ kHz \cite{RN32}. The CW-ODMR shows the linewidth broadening with strong MW fields. While the CW-ODMR linewidth reduces with the MW power, the linewidths of the pulsed-ODMR and the CE-pulsed-ODMR measurement stay unchanged when the $\pi$ pulse width exceeds $T_2^*$. Figure \ref{fig4}(a) also shows that pulsed-ODMR and CE-pulsed-ODMR are having almost the same linewidth. In order to determine whether the contrast reduction in CE-pulsed magnetometry degrades the sensitivity, we optimize the CW-ODMR sensitivity as a benchmark for comparison with the sensitivity of the CE-Ramsey measurement. We expect that CE-Ramsey combines the narrow linewidth with the advantage of a good SNR due to the continuous readout. All the following measurements are based on the near-optimal laser power for the CW-ODMR method, which can be easily achieved in the laboratory. Both the simulations and experimental results show that CE-Ramsey and regular Ramsey measurement yields similar results with the low-power laser \cite{Note}.

MW angular modulation is implemented as discussed in Sec. \ref{sec II}. Either frequency modulation (FM) or phase modulation (PM) can be applied. Taking the FM MW signal as the example, the signal can be described as
\begin{equation}
	S_{FM} = A \cos[2\pi f_0 t + \beta \sin (2\pi f_m t)]
	\label{eq3},
\end{equation}
where $f_0$ is the resonance frequency, $f_m$ is the modulation frequency, $\beta = f_d / f_m$ is defined as the modulation index in which $f_d$ is the frequency modulation amplitude. In the case of PM, the modulated phase term is $\phi_d \sin (2\pi f_m t)$, and the signal $S_{PM}$ can be expressed as (\ref{eq3}) by replacing $\beta$ with the phase modulation amplitude $\phi_d$. In applications, the FM signal is usually more precise than the PM signal due to the high frequency resolution of instrumentation. However, the PM signal provides a more stable phase reference, which is essential when applying a multi-frequency MW signal for the DR driving with lock-in detection. Another parameter that is worth discussing is the bandwidth of the modulated MW signal. The bandwidth of the modulated MW signal, i.e., Carson bandwidth, is defined as $BW_{FM} = 2(\beta + 1)f_m = 2(f_d+f_m)$, and the FM signal can be decomposed as
\begin{equation}
	S_{FM} = A \sum_{n=-\infty}^{\infty} J_n(\beta) \sin [2\pi (f_0 + n f_m) t]
	\label{eq4},
\end{equation}
where $J_n(\beta)$ is the Bessel function of the first kind. Figure \ref{fig4}(b) plots the maximum scalar factor of the lock-in ODMR spectra for $f_m$ = 1 kHz, 5 kHz, 10 kHz, and 20 kHz, and $f_d$ ranging from 1 kHz to 20 kHz. The main reason for the decrease of the measured scalar factors with $f_m$ is the bandwidth of the diamond magnetometer, as shown in Fig. \ref{fig3}(d). The lines are fitted according to the ODMR spectrum profile. The fitting only agrees well when $f_d+f_m$ is small but becomes more deficient with increasing $f_m$ or $f_d$ when $BW_{FM} >$ FWHM. Figure \ref{fig4}(c) plots the measured maximum scalar factors as a function of increasing $f_m$ with different $\phi_d=\beta$ to investigate the parameter dependence when applying the PM MW signal. The Carson bandwidth in the form of phase modulation is $BW_{PM}=2(\phi_d+1)f_m$. The scalar factor firstly rises because of the equivalent increasing $f_d = f_m\cdot \phi_d$. For larger modulation frequencies, the scalar factor is mainly limited by the magnetometer bandwidth. An additional reduction of the signal due to the modulated MW signal bandwidth can be found when $f_m$ is close to 20 kHz with $\beta = 4$, $BW_{FM} = 100$ kHz.

The CE-Ramsey scalar factor is also plotted in Fig. \ref{fig4}(c) for different modulation frequencies together with CW-ODMR data. It is compared with another CW-ODMR measurement, in which the MW signal bandwidth is fixed and $f_d = BW_{FM}/2 – f_m$ is changed according to $f_m$. The applied $BW_{FM}$ is 100 kHz to ensure the modulated fluorescence signal amplitude maximal. As depicted in Fig. \ref{fig2}(c), the signal measured by CE-Ramsey is demodulated with the reference set by the measurement cycle time. Since the effective fluorescence signal is majorly at the beginning of each cycle, the LIA output is small when the demodulation frequency is low and the repolarization time is long. Thus, the scalar factor of the CE-Ramsey measurement is small at the beginning and increases with the demodulation frequency. It also makes the CW-ODMR signal superior to the Ramsey signal with a $< 3$ kHz demodulation frequency. When the cycle time is 250 $\unit{\mu s}$, corresponding to a demodulation frequency of 4 kHz, the insufficient repolarization starts to reduce the contrast. Nevertheless, the CE-Ramsey scalar factor is still larger than the CW-ODMR over most of the investigated $f_m$. On the other hand, the high noise level from our instrumentation used in experiments at low frequencies, as shown see in Fig. \ref{fig4}(d), allows neither method to achieve the optimal signal-noise ratio.
One option for having a better sensitivity could be increasing the laser power to Watts. According to Fig. \ref{fig2}(b), the repolarization time is roughly 10 times shorter when the laser power meets k = 0.26. From Fig. \ref{fig3}(c), the gain of contrast is about 2 times. Taking the two factors into equation (\ref{eq2}), the shot-noise limit of the Ramsey measurement will be roughly 6 times lower. However, a Watt laser power could introduce more technical noise that might limit the sensitivity enhancement, which needs to be further investigated.

We investigate the instrument noise floor by measuring the noise spectrum and plotting the calculated root mean square voltages at different demodulation frequencies from 1 kHz to 20 kHz. When the setup operates with the laser switched on and the microwave signal off-resonant, the noise reduces by roughly 100 times from 1 kHz to around 10 kHz. However, the measured signals by each method reduce only by a few times, as shown in Fig. \ref{fig4}(c). In each measurement, the magnetometer outputs are acquired with the sampling rate of 900 Hz for 65536 pts, corresponding to a measurement time of 73 s. The minimum detected magnetic field is estimated by dividing the noise floor by the scalar factor. We plot the results calculated with these scalar factors in Fig. \ref{fig4}(d). It shows that the optimal demodulation frequency is at 9 kHz, with the measured magnetic field noise level at 2 pT. The sensitivity value is 17 $\unit{pT/\sqrt{Hz}}$ with 1 Hz normalization. The Ramsey measurement runs with ~4 MHz Rabi frequency and time interval $\tau_m = 6.42 \unit{\mu s}$. Meanwhile, according to the scalar factors measured in Fig. \ref{fig4}(c), the optimal CW-ODMR signal is only 80\% of the signal measured by the CE-Ramsey method, and the optimized noise level plotted in Fig. \ref{fig4}(d) is roughly $2.5 \unit{pT}$. The values are measured and calculated with MW driving the $\ket{0} \rightarrow \ket{-1}$ transition, hyperfine lines included. Considering the sensitivity enhancement factor of 1.3 times with DR driving, we expect the CW-ODMR method also measures a 2 pT noise level. In the magnetometer intrinsic noise measurements, batteries are used to generate a bias field with the coils to keep the low magnetic field noise in the shields. However, the continuously discharging of the batteries induces bias-field reduction during the measurements. The continuous shifting of resonant frequency makes it difficult to run the Ramsey measurements, which require on-resonant MW pulses. The CW-ODMR method is used for characterizing the intrinsic noise of the setup in this work. The bias field drift can be tracked by repeatedly sweeping the MW frequency. In order to keep the bandwidth of the modulated MW signal approaching the ODMR linewidth in the measurement, the applied parameters are $f_m$ = 9 kHz, $\phi_d$ = 2, corresponding to $f_d$ = 18 kHz and BW = 54 kHz instead of the 100 kHz bandwidth used in Fig. \ref{fig3}(c).

\begin{figure}[t]
	\includegraphics[scale=1.0]{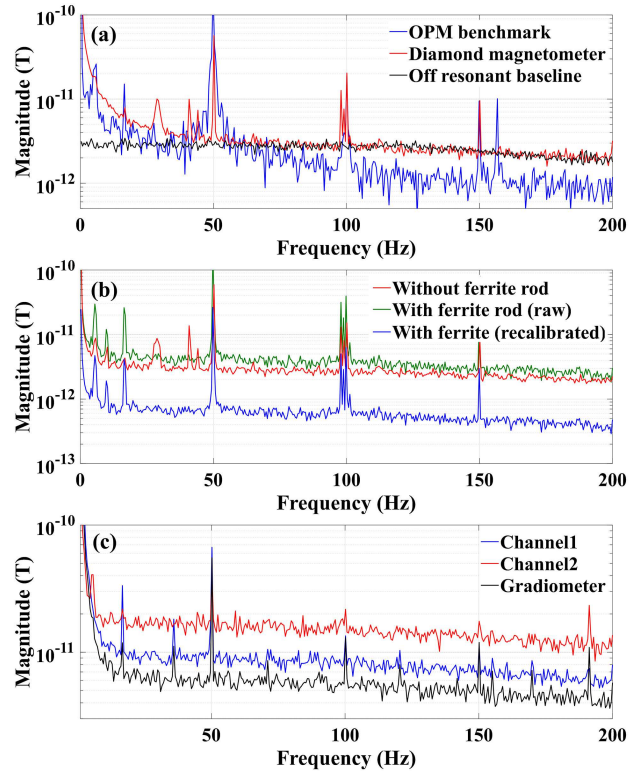}
	\caption{\label{fig5} 
		(a) Magnetic field noise spectrum measured by CW-ODMR method with hyperfine driving and DR driving, compared to the spectrum measured by the OPM sensor. The black line is the off-resonant baseline showing the intrinsic noise of the diamond magnetometer. With the measurement time of 73 s, all of the plotted spectra in this figure have the same frequency resolution of 13.7 mHz. The magnitudes are plotted in logarithm, so are the following two figures.
		(b) The green line and red line are the noise spectra analyzed from the direct outputs of measurements with and without the FG. The blue line shows the measured magnetic field noise calibrated by the battery discharging signal difference of the two measurements. In order to see the low-frequency noise spikes, the discharging ramp of the output is removed by linear fitting before the spectral analysis. 
		(c) Noise spectra measured by the gradiometer setup: the probe channel 1 (blue), reference channel 2 (red), and gradiometry signal (black) are measured, respectively. The spectrum of each single channel is measured by applying single resonance driving, and the spectrum of the gradiometer is measured by combining the two channels and operated with the DR driving scheme.}
\end{figure}

The resonantly measured magnetic field noise spectrum is depicted in Fig. \ref{fig5}(a). The spectrum is calculated by the Fourier transform of the Lock-in amplifier output, dividing the measured scalar factor and the gyromagnetic ratio. The magnetometer has also been calibrated by applying known fields modulated at 182 kHz \cite{Note}. The magnetic field noise inside the shielding is measured by the OPM (Quspin Inc., QTFM) as a benchmark with the nominally 1 $\unit{pT/\sqrt{Hz}}$ sensitivity. The diamond magnetometer measures the magnetic field noise with about 10 G bias field generated by the coils powered by batteries. The coils are turned off when the OPM measures the noise to ensure the required bias field range. The batteries ensure the coil noise lower than the residual magnetic noise inside the shields so that the OPM and the diamond magnetometer could measure the same magnetic field noise floor. The diamond magnetometer measures a noise floor of $2 - 3$ pT for frequencies larger than 50 Hz. The noise spectrum at a frequency lower than 50 Hz is partly limited by the residual low-frequency noise inside the shields according to the spectrum measured by the OPM. On the other hand, the discharging of the batteries is another reason for the higher noise level of the diamond magnetometer at low frequencies. The discharging can be removed by a linear fitting so that low-frequency noises can be well characterized. In Fig. \ref{fig5}(b), the spectra are calculated and plotted with the discharging signal removed, and the low-frequency noise spikes are better characterized than the spectrum in Fig. \ref{fig5}(a). The off-resonant baseline in Fig. \ref{fig5}(a) marks the intrinsic noise level of the diamond magnetometer, which is $2 - 3$ pT from 0 to 200 Hz. The noise level is consistent with the result in Fig. \ref{fig4}(d), which is limited by the optical noise, showing that we measure a shot-noise limited sensitivity of $17 \unit{pT/\sqrt{Hz}}$.

The most efficient way to overcome the sensitivity limit is using a flux concentrator (FC). The increase in sensitivity using a FC follows a simple geometric consideration, i.e., the ratio of the areas at the surface proximal to the diamond vs. the distal surface defines the amplification \cite{RN29}. It is also possible to use a ferrite rod with a millimeter diameter as the FG \cite{RN36} so that the diamond can measure a remote magnetic field with concentrated flux. Although the ferrite rod cannot have the same flux amplification factor as the FC, the flux concentration is still remarkable due to the high magnetic permeability of the ferrite material, and the FG holds the spatial resolution. Experimentally, the magnetic field generated by the coil is guided and concentrated through the rod to the diamond. The magnetic field signal induced by the battery discharging is detected as the calibration signal. The discharging signal is a linearly decreasing field, and the slopes measured by magnetometer with and without the FG are compared. With the FG, the magnetic field signal decreases 6.3 times faster than the measurement without FG \cite{Note}.
It indicates that the flux is amplified by a factor of 6.3 times. Figure \ref{fig5}(b) shows the comparison of the detected magnetic field noise floor. In order to have a better analysis of the low-frequency magnetic field noise, the discharging signal is removed from the signal traces by linear fitting before performing the Fourier transform. Then, spikes with frequency under 20 Hz can be seen in the spectrum, and the signals come from the cooling fan motors (typically 1000 r/min, roughly 17 Hz) of equipment outside the shields. There is a window on the shielding cube in the direction of the FG. The magnetic noise spikes are guided through the ferrite to the diamond in the measurement. The noise floor translated from the raw output of measurement with the FG is read as 3 pT. With the 6.3 times flux gain, the noise floor measured with the FG is estimated between $0.3 - 0.7$ pT over the bandwidth of $20 - 200$ Hz, as the spectrum shown in Fig \ref{fig5}(b). The 1 Hz normalized magnetic field noise spectral density is $2.6 - 6 \unit{pT/\sqrt{Hz}}$.

Using the gradiometer configuration described in Fig. \ref{fig1}(b), we measure the diamond magnetic gradient field noise as shown in Fig. \ref{fig5}(c). Respectively, the MW signal is applied to drive the transition of $\ket{0}\rightarrow\ket{+1}$ in the probe channel (Channel 1) and $\ket{0}\rightarrow\ket{-1}$ in the reference channel (Channel 2). The probe channel measures a 5 pT noise level due to the driving of a single transition. The noise level of channel 2 is $10 - 20$ pT, mainly because of the MW inhomogeneity generated by the loop antenna. Nevertheless, by combing the two signals as the DR driving scheme shown in Fig. \ref{fig2}(a), the gradiometer suppresses the common-mode line shifts, and the differential signal is amplified. Each gradiometer channel can also operate with DR driving, for which four groups of MW frequencies are required. The differential signal can still be amplified by applying MW signals with a $180\deg$ phase difference regarding the two channels. The differential signal gain could be twice larger because the measurement is operated in two separate diamonds. In principle, the gradient field noise increases by $\sqrt 2$ compared to a single-channel magnetometer. Thus, the sensitivity of the differential signal could be $\sqrt 2$ times better. In the measurement with single resonance driving in each channel, the measured gradiometer intrinsic noise is $4 - 6$ pT in the frequency range of $20 - 200$ Hz and a channel distance of roughly 10 cm. 

\section{Conclusions and outlooks}

This work demonstrates the diamond DC magnetometry and gradiometry with low optical pumping power. Fields ranging from few picotesla to subpicotesla level are measured within a sensing volume of 0.125 $\unit{mm^3}$. We take advantage of the narrow linewidth and the low noise floor of the lock-in detection technique, using the Ramsey method with low power continuous excitation. With the optimized parameters, the CE-Ramsey measurement can achieve a better sensitivity than CW-ODMR measurement. The real field measurements indicate the intrinsic magnetic noise level of the diamond magnetometer is $2 - 3$ pT with a 13.7 mHz frequency resolution, corresponding to a bandwidth normalized noise spectral density of $17 \unit{pT/\sqrt{Hz}}$. The gradiometry measurements also demonstrate a $<10$ pT gradient field noise with the same frequency resolution. By applying a ferrite rod with a 2 mm end diameter as the flux guide, the noise level (spectrum calibrated by the battery discharging signal) reduces to  $0.3 - 0.7$ pT, corresponding to a bandwidth normalized noise spectral density of $2.6 - 6 \unit{pT/\sqrt{Hz}}$. For practical applications, one could combine the FG and the gradiometer to take the combined advantage of the sensitivity, spatial resolution, and magnetic field noise suppression. The FG could be further optimized for higher flux gain by changing its geometry and structure, such as the gap distance between the diamond and the FG end and the structure with a flux return ferrite. Moreover, upon compromising the spatial resolution, the setup could yield a magnetic field sensitivity of a few hundred $\unit{fT/\sqrt{Hz}}$ at ambient environment if the flux concentrator geometry is used. Techniques such as double quantum magnetometry\cite{RN40}, spin bath driving, multiple NV orientation driving, and close-loop measurements could advance the sensitivity and stability \cite{RN37,RN38}. With the demonstrated intrinsic noise level and spatial resolution, diamond magnetometer can be at the forefront for a wide range of sensing applications in classical and quantum technologies.

\section{Acknowledgments}
We thank Axel Griesmaier for offering excellent photodetectors. We also thank Andrej Denisenko, Jianpei Geng, Durga Dasari, and Ayman Mohamed for fruitful discussions. We acknowledge financial support by EU via the project ASTERIQS, and the ERC Advanced Grant No. 742610, SMel, and the BMBF via the project MiLiQuant and the DFG via the GRK 2198 and 2642. In addition, we acknowledge the support of the Japan Society for the Promotion of Science (JSPS) KAKENHI (No. 17H02751).

\nocite{*}


%

\clearpage
\section*{Supplemental information: Diamond magnetometry and gradiometry towards subpicotesla DC field measurement}
\renewcommand{\theequation}{S\arabic{equation}} 
\renewcommand{\figurename}{S-FIG.} 
\setcounter{figure}{0}
\setcounter{equation}{0}
\subsection{Setup details}
The diamond samples are (111)-oriented (0.5 mm)$^3$ cubes obtained by laser-cutting and polishing from a 99.97\% $^{12}$C enriched single crystal. The crystal is grown by the temperature gradient method at high pressure and high temperature (HPHT) conditions. The initial nitrogen concentration of the original crystal was 1.4 ppm. After irradiation of 2 MeV electrons and annealing (1000 $^\circ$C for 2 h in vacuum), the NV concentration reached ~0.4 ppm.

The laser used in experiments is the low noise version of the Lighthouse Sprout-G 532 nm laser. The home-built photodetector is based on the Hamamatsu S3590-09 model with an active area of 10×10 mm$^2$ and a simple trans-impedance amplifier circuit with a gain of 5.1 kV/A. The detected fluorescence and laser reference signals directly go to the differential input of the lock-in amplifier (LIA, Zurich Instruments, HF2LI). Two vector signal generators (Rohde \& Schwarz, SMIQ03B) are used to generate MWs for the double resonance (DR) driving. In the CW-ODMR measurements, the MWs are mixed with a PM modulated RF signal generated by a function generator (Rigol DG1022), of which the second channel generates a 2.163 MHz signal mixed with the MW by a mixer (Minicircuits, ZAM-42) to drive the hyperfine lines. In the MW pulsed measurements, the MW signal is gated by a data timing generator (Tektronix, DTG5274) and a switch (Minicircuits, ZASWA-50DR+). The combined multi-frequencies MW signal is amplified by the Amplifier-Research 50S1G4 and finally applied to the diamond through the dielectric resonator antenna \cite{RN1}. The bias field is generated by the home-built 3D coil system, which are powered by 12 V lead-acid batteries (XCell XP 18-12) in the noise floor measurements to keep the low magnetic field noise level. As illustrated in Fig. 1(a), the setup, including the coils, is installed in the magnetic shields. The shields consist of an inner $\mu$-metal cube and an outer aluminum cube. In the magnetic field measurement with the flux guide, the flux guide is machined from a MN60 ferrite rod, with the permeability of 6500. All the measurements acquire signals with sampling rate of 900 Hz for 65536 pts, corresponding to about 73 s.

\subsection{Operation protocols}
\subsubsection{CW-ODMR operation}
The CW-ODMR operation has been described in several publications \cite{RN2,RN3,RN4}. In the main text, we have discussed the optimization of operation parameters theoretically and experimentally. In order to determine the scalar factor when using double resonance driving, the two MW signals are swept in opposite directions centering at the two resonant frequencies with the same span. The slope of the ODMR spectrum derivative at the resonant frequency is denoted as the scalar factor $k_s$. The magnetic field is then measured by applying the two resonant MW frequencies synchronously, and the magnetic field is recorded as the output voltage level dividing the scalar factor.

\subsubsection{CE-Ramsey operation}
\begin{figure}[htb]
	\includegraphics{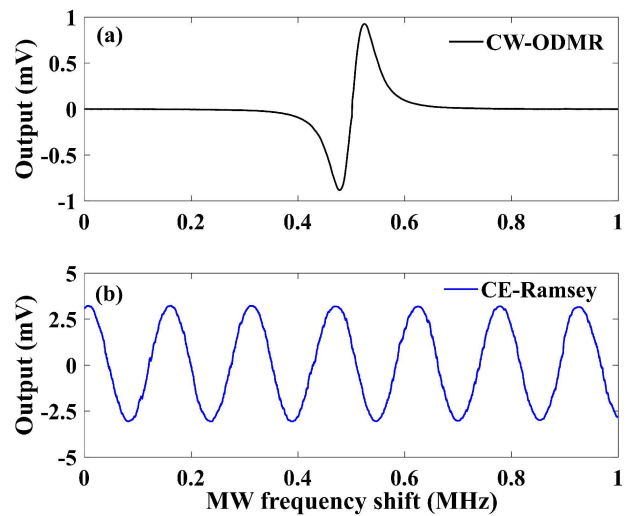}
	\caption{\label{sfig1}Magnetometer response to different MW frequency offsets measured by (a) CW-ODMR method; (b) CE-Ramsey method.}
\end{figure}

The CE-Ramsey operation is based on the sequence described in Fig. 2(b). The magnetic field sensing time $\tau_m = 6.42 \unit{\mu s}$ is chosen according to the optimization of max: $C(\tau_m)\cdot\tau_m$. The repolarization time $\tau_r$ is calculated according to the optimized demodulation frequency as
\begin{equation}
	\tau_r=(T_{seq}-3\pi/\Omega_R)/2-\tau_m
	\label{s1}.
\end{equation}
The external magnetic field is measured by continuously running the sequence with the fixed $\tau_m$. In order to suppress the thermal fluctuation, the double quantum magnetometry can also run with the CE-Ramsey scheme as well by replacing the pulses with double quantum magnetometry pulses \cite{RN5}. The scalar factor of the magnetometer is determined by sweeping the MW frequency as the equivalence of magnetic field changes. S-Figure.\ref{sfig1} shows the magnetometer response to different MW frequency offsets. Both of the results are measured with single resonance driving for comparison.

\subsection{Simulation model}
The transition kinetics simulation follows the energy level diagram described in Fig. 1(d). In the diagram, the charge state conversion is neglected because the NV$^0$ has no contribution to the valid signal. Another reason is that the low pumping rate preserves the major population on NV$^-$ states \cite{RN6,RN7}, and the continuous excitation maintains the population of NV$^-$ and NV$^0$ stable. The simulation includes the green laser excitation, T$_1$ decay, and the MW driving in the single quantum (SQ) basis of the ground state. The MW driving model is based on the Bloch equations \cite{RN8}:
\begin{eqnarray}
	&\displaystyle{\frac{\partial\rho_{00}}{\partial t}}& = \Gamma_p\rho_{00}-\Gamma_1(\rho_{00}-\rho_{11})+\frac{i\Omega_R}{2}(\rho_{01}-\rho_{10})
	\label{s2},\\
	&\displaystyle{\frac{\partial\rho_{01}}{\partial t}}& = (-\Gamma_2+i\Delta)\rho_{01}+\frac{i\Omega_R}{2}(\rho_{00}-\rho_{11})
	\label{s3}.
\end{eqnarray}
in which $\rho_{ij}$ are elements of the density matrix, $\Gamma_p$ is the optical pumping rate, $\Gamma_1 = 1/T_1, \Gamma_2^* = 1/T_2^*$ , and $\Delta$ is the MW offset frequency. Here we take the $\ket{0}\rightarrow\ket{+1}$ basis. Combined (\ref{s2}), (\ref{s3}) with the optical pumping kinetics, the model is described as:
\begin{widetext}
	\begin{eqnarray}
		&\displaystyle{\frac{\diff{n_1}}{\diff t}}=-\Gamma_p n_1+R_{fl}n_4+R_{81}n_8
		-\frac{1}{3}\Gamma_1(2n_1-n_2-n_3)-\Omega_R\mathrm{Im}(\rho_{01})
		\label{s4},&\\
		&\displaystyle{\frac{\diff{n_2}}{\diff t}}=-\Gamma_p n_2+R_{fl}n_5+R_{82}n_8-\frac{1}{3}\Gamma_1(2n_2-n_1-n_3)\label{s5},&\\
		&\displaystyle{\frac{\diff{n_3}}{\diff t}}=-\Gamma_p n_3+R_{fl}n_6+R_{83}n_8-\frac{1}{3}\Gamma_1(2n_3-n_1-n_2)-\Omega_R\mathrm{Im}(\rho_{01})\label{s6},&\\
		&\displaystyle{\frac{\diff{\mathrm{Re}(\rho_{01})}}{\diff t}} = \Delta\cdot\mathrm{Im}(\rho_{01})-\Gamma_2^*\mathrm{Re}(\rho_{01})
		\label{s7},&\\
		&\displaystyle{\frac{\diff{\mathrm{Im}(\rho_{01})}}{\diff t}} = \Delta\cdot\mathrm{Re}(\rho_{01})-\Gamma_2^*\mathrm{Im}(\rho_{01})-\frac{\Omega_R}{3}(n_3-n_1)\label{s8},&\\
		&\displaystyle{\frac{\diff{n_4}}{\diff t}}= -R_{fl}n_4+\Gamma_pn_1-R_{47}n_4
		\label{s9},&\\
		&\displaystyle{\frac{\diff{n_5}}{\diff t}}= -R_{fl}n_5+\Gamma_pn_2-R_{57}n_5
		\label{s10},&\\
		&\displaystyle{\frac{\diff{n_6}}{\diff t}}= -R_{fl}n_6+\Gamma_pn_3-R_{67}n_6
		\label{s11},&\\
		&\displaystyle{\frac{\diff{n_7}}{\diff t}}=R_{47}n_4+R_{57}n_5+R_{67}n_6-R_{78}n_7
		\label{s12},&\\
		&\displaystyle{\frac{\diff{n_8}}{\diff t}}=-R_{78}n_7-(R_{81}+R_{82}+R_{83})n_8
		\label{s13},&
	\end{eqnarray}
\end{widetext}
In the model (\ref{s4}) - (\ref{s13}), $n_i$ is the population of state $i$, $R_{ij}$ is the transition rate from state $i$ to state $j$, and $R_{fl}$ = 66 MHz is the fluorescence emission rate. The intersystem crossing parameters $R_{82} = R_{83} =0.7$ MHz, $R_{81} = 1$ MHz, $R_{57} = R_{67} = 53$ MHz, and the metastable states emission rate $R_{78}$ = 1000 MHz \cite{RN9,RN10}. The fluorescence signal is taken as the sum of the excited states population, i.e., $n_4+n_5+n_6$. The background is taken as the signal when the spins are fully initialized without MW. To calculate the bandwidth of the magnetometer, we applied the modulated MW signal with frequency shift follows $\Delta(t, f_m, f_d)$, where $f_m$ is the modulation frequency, $f_d$ is the frequency deviation, and the modulation index $\beta=f_d/f_m=\phi_d$.

\subsection{Continuous excitation}
\begin{figure*}[t]
	\includegraphics[scale=1]{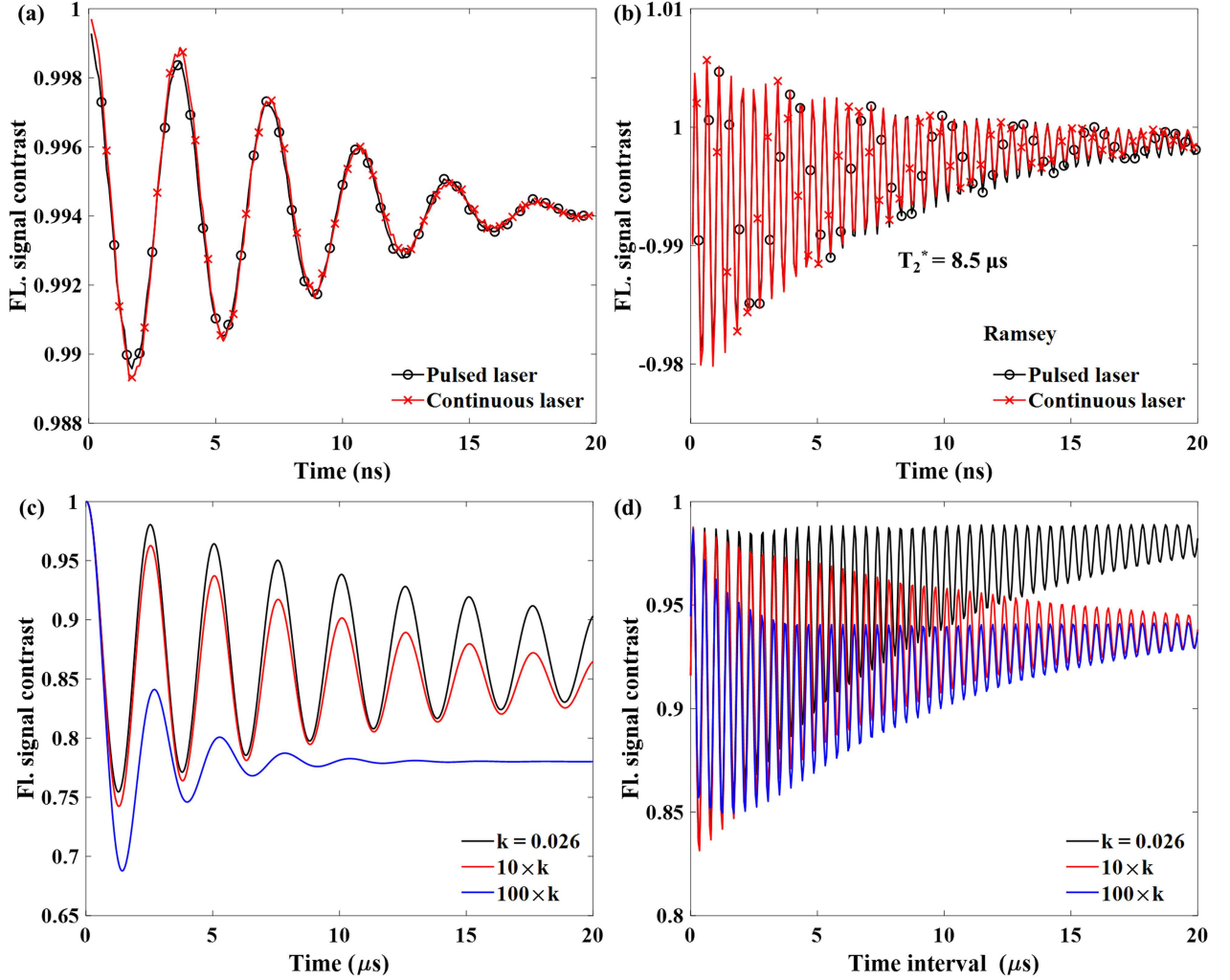}
	\caption{\label{sfig2} 
		(a) Rabi oscillations measured by applying continuous laser compared with regular Rabi measurement using pulsed laser.
		(b) Results comparison of CE-Ramsey measurement and the normal Ramsey measurement.
		(c) Simulated results of CE-Rabi measurements with increasing optical pumping rate. The pumping rate $k = 0.026$ corresponds to the laser power of 80 mW. 
		(d) FID signals calculated with CE-Ramsey sequence with increasing optical pumping rate, and $\Omega_R$ = 2.5 MHz.}
\end{figure*}
One important question is, what would be the cost of using continuous excitation for getting away from the technique noises. Experimentally, Rabi oscillation and the free induced decay (FID) signal are measured with both pulsed laser and continuous laser excitation. Fluorescence signals are all acquired with a $10 \unit{\mu s}$ gate after the MW pulses so that the calculated contrast of the two measurements can be comparative. As shown in S-Fig. \ref{sfig2}(a) and (b), the two curves in each figure are hardly distinguishable. The influence of continuous excitation with low laser power can be neglected. In order to investigate the continuous excitation effects on interferometry measurements, we simulate the process with the model described above. According to the Bloch equations (\ref{s2}) and (\ref{s3}), the repolarization has an equivalent effect as $T_1$, that it induces Rabi oscillation signal decay in the duration of MW pulse. S-Figure \ref{sfig2}(c) plots the simulated Rabi oscillations with increasing laser pumping rate. The established Rabi frequency $\Omega_R$ = 2.5 MHz. The signal decay becomes faster with a higher pumping rate. The average contrast level decreases with increasing pumping rate because the metastable state accumulates more population with appearance of MW when the pumping power is higher. With the presence of $\Gamma_p$ and $\Gamma_1$, the Rabi flopping frequency is given as
\begin{equation}
	\Omega_\Gamma=\sqrt{\Omega_R^2-(\frac{\Gamma_p}{4}-\frac{\Gamma_1}{2})^2}
	\label{s14}.
\end{equation}
In the experiments, $\Omega_R\gg\Gamma_p, \Gamma_1$ , and there would be no significant Rabi frequency change until $\Gamma_p\approx\Omega_R$. In any case, the pulse width error in Ramsey sequence can be experimentally compensated. According to (\ref{s3}), the repolarization does not contribute to the dephasing in the kinetics during the phase integration time interval $\tau_m$. S-Figure \ref{sfig2}(d) shows the simulated FID signal of CE-Ramsey measurements with increasing pumping rate. In the simulation, the MW frequency is on-resonant to the hyperfine line of I = 0, and the oscillation comes from the $^{14}$N nuclear spin I = $\pm$1 precessions at $\pm$2.2 MHz. The result shows a faster decay of contrast when the pumping rate is higher. It is because that the repolarization accumulates during $\tau_m$ due to the continuous pumping. 

Techniques such as balancing-detector, heat sinking can also be applied for better SNR and stability. While these techniques cannot address all the issues in regular Ramsey measurements, continuous excitation can fix all the problems as mentioned in the main text. For example, a balancing-detector cannot eliminate pulsed readout when the two arms are not well balanced, but by continuous excitation there will be no background fluorescence pulses. Heat sinking cannot prevent the thermal dynamics over the diamond induced by laser pulses, and it could be a reason for linewidth broadening in bulk diamond measurements. Combining continuous excitation with a heat sinking can suppress most of the thermal fluctuations, and thermal stability can be further improved by monitoring the laser power fluctuations. In conclusion, continuous excitation using a low pumping power is beneficial with minimal influence on the signal contrast.

\subsection{Linewidth of the diamond sample}
Our diamond sample has NV ensembles with a long dephasing time of $T_2^* = 8.5 \unit{\mu s}$, and it ensures the very narrow ODMR linewidth. The CW-ODMR linewidth of the diamond can be estimated by \cite{RN8}
\begin{equation}
	\Delta\nu=\frac{1}{2\pi}\sqrt{\Gamma_2^2+\frac{\Omega_R^2\Gamma_2}{2\Gamma_1+\Gamma_p}}
	\label{s15},
\end{equation}
where $\Gamma_2=\Gamma_2^*+\Gamma_c$, and $\Gamma_c$ is the optical cycles. When both MW power and laser power approach 0, the CW-ODMR linewidth limit can be estimated as $\Gamma_2^*/2\pi\approx19$ kHz. The minimum linewidth detected experimentally is 28 kHz. It was difficult to measure the ODMR linewidth with further lower MW power because of the extremely low contrast and the linewidth broadening due to the magnetic field drifts during a long averaging time.

On the other hand, pulsed-ODMR reaches the linewidth limit when the $\pi$-pulse time $T_\pi\approx T_2^*$. The linewidth is estimated as $\Delta\nu=(2\sqrt{\ln 2})/(\pi T_2^* )\approx62$ kHz \cite{RN8}, while experimentally the minimum linewidth is 85 kHz. One possible reason for the linewidth broadening is that the $\pi$-pulse with a width exceeding T2* becomes inaccurate due to the dephasing limit. In the experiment, we can only estimate the $\pi$ pulse width by fitting Rabi frequency to the MW power. The contrast becomes poor for further measurements also with longer $\pi$-pulse width under lower MW power. 

\subsection{\label{F}Signal contrast}
\subsubsection{CW-ODMR contrast}
This section discusses the contrast reduction induced by the MW modulation in magnetic field measurements. Figure 3(d) in the main text shows that the magnetometer bandwidth has a dependency on the optical pumping rate. MW modulation in the CW-ODMR measurement scheme directly leads to a response of signal contrast following the figure. Experimentally, the MW modulation parameters are $f_m$ = 9 kHz, $f_d=\phi_d f_m = 18$ kHz. Thus, the differential signal $\diff S(f) = S(f-f_d )-S(f+f_d )$ can be calculated basing on the regular CW-ODMR spectrum measured without MW modulation, and $S(f)$ is the detected fluorescence signal. The peak-peak amplitude of the differential signal $\diff S(f)$ is 17.1 mV, while the peak-peak amplitude of the lock-in detected ODMR signal with the 9 kHz MW modulation is 1.81 mV according to S-Fig. \ref{sfig1}(a). Thus, the reduction of the CW-ODMR contrast due to MW modulation can be calculated as 9.45 times. The simulated frequency response of the diamond magnetometer plotted in Fig. 3(d) also indicates a similar contrast reduction.
\begin{figure*}[t]
	\includegraphics[scale=1]{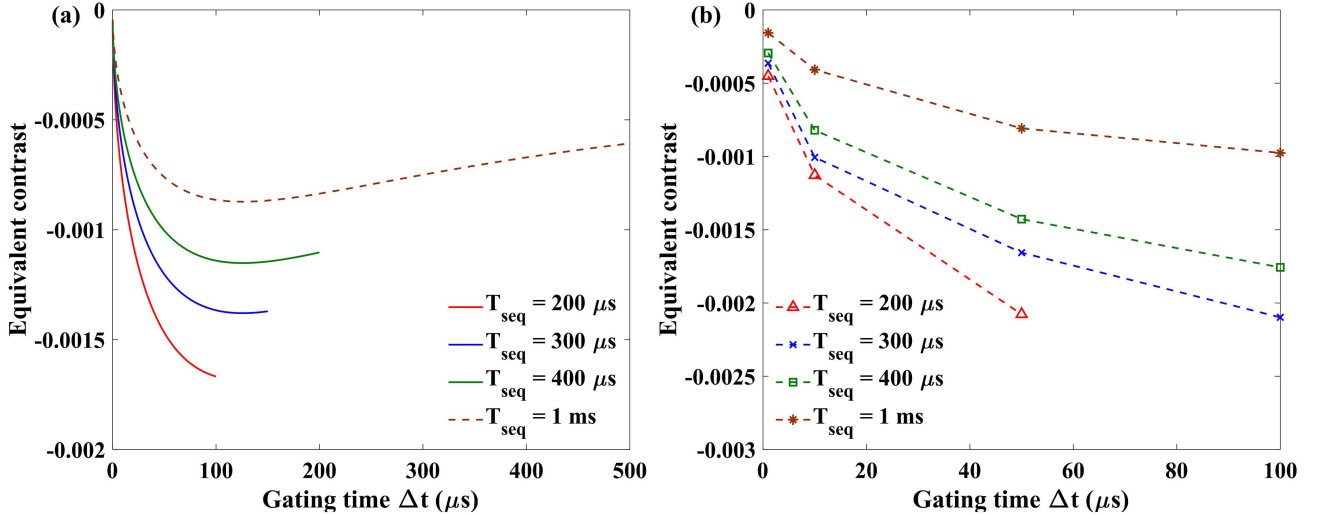}
	\caption{\label{sfig3} 
		(a) is the calculated equivalent contrast with different acquisition gating time and the sequence cycle time. 
		(b) is the measured $C_{eqv}$ in experiments.}
\end{figure*}
\subsubsection{Ramsey contrast}
As for the Ramsey measurement, the main reason for the contrast reduction is the insufficient repolarization due to the short cycle time in the lock-in detection, which is discussed in Fig. 2(c). It is shown in the figure that the spins can mostly be repolarized by the 80 mW laser in 0.5 ms. The recovery curve is fitted by $U_{fl}=-A_{fl} \exp(-t/\tau_{fl})$, where $A_{fl}$ = 105 mV is the largest fluorescence decrease, $\tau_{fl}$ = 0.1 ms is the time constant of the signal decay. When the cycle time is short, the spins cannot be fully polarized. In order to estimate the output of the LIA, we can choose a reference time $t_0$ and get
\begin{widetext}
	\begin{equation}
		U_{LIA} = \left[\int_{t_0-\frac{T_{seq}}{2}}^{t_0} U_{fl} \diff t-\int_{t_0}^{t_0+\frac{T_{seq}}{2}}U_{fl}\diff t\right] \left(\frac{T_{seq}}{2}\right)^{-1} \exp\left(-\frac{\tau_m}{T_2^*}\right)
		\label{s16},
	\end{equation}
\end{widetext}
where $\tau_m = 6.42 \unit{\mu s}$, is the field measurement time. With demodulation frequency of 9 kHz, the cycle time $T_{seq}\approx 110 \unit{\mu s}$, and the reference time t$_0$ is taken as 150 $\unit \mu$s. It can be calculated that $U_{LIA} = 6.2$ mV. Meanwhile, according to S-Fig. \ref{sfig1}(b), the experimentally detected signal $U_{LIA}'=U_0-U_{-1}\approx6.3$ mV, where $U_0$ and $U_{-1}$ are the fluorescence signal at the corresponding spin states. It can be concluded that the insufficient repolarization is the reason for contrast reduction in the Ramsey measurement using lock-in detection. The Ramsey contrast is calculated as $C_{det}=U_{LIA}'/(U_0+U_{-1}) \approx0.17\%$, where the average fluorescence signal $(U_0+U_{-1})/2$ is estimated as 1.88 V according to the experimental data.
\subsubsection{Continuous readout and gated readout}
One of the commonly used methods to increase signal contrast in NV interferometry measurements is to gate the acquired fluorescence pulses. Intuitively, gated acquisition can avoid the invalid fluorescence detection induced by repolarization. In the FID measurements of S-Fig. \ref{sfig2}(b), the gated readout is acquired for comparison. Each sequence time is $T_{seq} = 200 \unit{\mu s}$, and the gating time is 10 $\unit\mu$s. The detected signal magnitude is estimated by
\begin{widetext}
	\begin{equation}
		U_{G} = \left[\int_{t_0-\frac{T_{seq}}{2}}^{t_0-\frac{T_{seq}}{2}+\Delta t} U_{fl} \diff t-\int_{t_0}^{t_0+\Delta t}U_{fl}\diff t\right] \left(\Delta t\right)^{-1} \exp\left(-\frac{\tau_m}{T_2^*}\right)
		\label{s17}.
	\end{equation}
\end{widetext}
When $\tau_m$ = 0, $U_G$ = 38.3 mV with the repolarization reference time $t_0$ = 150 $\unit\mu$s, and the signal contrast is calculated as $U_G/U_0\approx2\%$. On the other hand, the gated acquisition deteriorates the sensitivity by a factor of $\sqrt{T_{seq}/\Delta t}$ (similar sensitivity deterioration happens in Ref. \cite{RN11}, in which the equation 3 mistakenly writes the factor as $\sqrt{\Delta t/T_{seq}}$). Another factor of $\sqrt{T_{seq}}$ is multiplied for consideration of the magnetic field sampling rate. Therefore, we use the equivalent contrast parameter defined as $C_{eqv}=C_{det}/\sqrt{(T_{seq}/\Delta t)}/\sqrt{(T_{seq}/T_{ref}}$ for the investigation. $C_{eqv}$ includes both the detected contrast and the sequence parameters to represent the sensitivity changes with different gating time and sequence cycle time. $T_{ref}=110 \unit{\mu s}$ is used for comparison with the $f_m=9$ kHz measurement in this work. The gated readout is equivalent to the LIA readout when $\Delta t=T_{seq}/2$. 

S-Figure \ref{sfig3}(a) plots the calculated $C_{eqv}$ with different gating time and sequence cycle time. The fluorescence reference time is chosen as $t_0=T_{seq}/2+50 \unit{\mu s}$ to represent the insufficient repolarization when $T_{seq}$ is short. Only with a long enough cycle time, there is an optimized gating time around 0.1 ms. Considering the sensitivity deterioration induced by the sequenced sampling, the acquisition with $\Delta t=T_{seq}/2$ and short $T_{seq}$, i.e. the lock-in acquisition used in this work, is superior. The experimentally measured contrast is plotted in S-Fig \ref{sfig3}(b). The measured results show the same trends as the calculation. The difference between the measured and the calculated contrast could result from the minor difference in fluorescence reference time. In conclusion, the results show that the continuous readout with a LIA is superior in sensitivity compared to the gated readout in the case of low laser power pumping.

\subsection{Sensitivity estimation}
\subsubsection{CW-ODMR}
In order to estimate the shot noise limited sensitivity, the linewidth, contrast, and photon detection rate have to be determined experimentally. According to Fig. 3(c), the measured $\nu/C_{det}$ with hyperfine lines driving and DR driving is $2.37\times10^6$ Hz. To estimate the photon detection rate $R$, we measured the photodetector output level $U_{PD}$ = 3.76 V with 1 $\unit{M\Omega}$ termination. Thus, the detected photon count rate is
\begin{equation}
	R=\frac{U_{PD}}{G_{TI}\cdot q}=4.6\times10^{15} \unit{Hz}
	\label{s18},
\end{equation}
where $G_{TI}$ = 5100 is the transimpedence gain, and $q$ is the electron charge. Therefore, the shot noise limited sensitivity can be estimated according to equation (2) as 0.96 pT/Hz$^{1/2}$. However, this sensitivity limit is calculated without considering the contrast reduction due to the MW modulation. The parameter $\nu/C_{det}$ should be corrected with a factor depending on the MW modulation frequency. According to the discussion of the CW-ODMR contrast in Section \ref{F}, the factor 9.45 can be considered in the estimation. As a result, the shot noise limited sensitivity of CW-ODMR measurement with 9 kHz MW modulation is estimated as 9.1 pT/Hz$^{1/2}$.

Experimentally, the scalar factor measured by CW-ODMR with 9 kHz MW modulation is calculated according to S-FIG. \ref{sfig1}(a) as $\diff S/\diff f\cdot\gamma_{NV}=2.97\times 10^{-6} \unit{V/nT}$. In the slope measurement, only hyperfine lines driving is used in the CW-ODMR measurement in order to make the result comparable to the result of single-quantum Ramsey magnetometry measurement. The measured shot noise level at the demodulation frequency 9 kHz is 7.2 nV in the 73 s measurement, corresponding to 61.5 nV/Hz$^{1/2}$ with bandwidth normalized to 1 Hz. Thus, the sensitivity can be calculated as 20.7 pT/Hz$^{1/2}$. Considering the enhancement factor of about 1.3 times with DR driving, we can expect a sensitivity of 15.9 pT/Hz$^{1/2}$ in CW-ODMR measurements. Since we illuminate the entire diamond with volume of (0.5 mm)$^3$, volume-normalized sensitivity can be calculated as $\eta_v=\eta\sqrt{V}$, which is 5.6 $\unit{pT/(Hz\cdot mm^{-3})^{1/2}}$.

\subsubsection{Ramsey measurement}
As for the CE-Ramsey measurement, the dephasing time $T_2^* = 8.5 \unit{\mu s}$ is measured as shown in S-Fig. \ref{sfig2}(b). According to the discussion in Section \ref{F}, the Ramsey contrast with sequence time of $T_{seq} \approx 110 \unit{\mu s}$ is 0.17\%. The collected photon per measurement is calculated based on the photon detection rate. Thus, Eq. (1) in the main text is simplified as
\begin{equation}
	\delta B_{Ramsey}=\frac{\hbar}{g\mu_B}\frac{1}{C_{det}\tau_m\sqrt{Rt}}
	\label{s19},
\end{equation}
With the 1 s measurement time, the sensitivity $\eta=\delta B\sqrt t$ is calculated to be 7.7 pT/Hz$^{1/2}$. The sensitivity is close but superior to the CW-ODMR sensitivity even without considering the double quantum magnetometry.

Experimentally, the scalar factor measured by the CE-Ramsey method is calculated according to S-Fig. \ref{sfig1}(b), from which the maximum slope is $3.6\times10^{-6}$ V/nT. With the shot noise demodulated at 9 kHz, the sensitivity of the Ramsey measurement is calculated as 17.1 pT/Hz$^{1/2}$. There could be an enhancement if double quantum magnetometry is applied. Then, the sensitivity should be better than the value measured by the CW-ODMR method. 

\subsection{Calibration of the magnetometer}
\begin{figure}[ht]
	\includegraphics[scale=1]{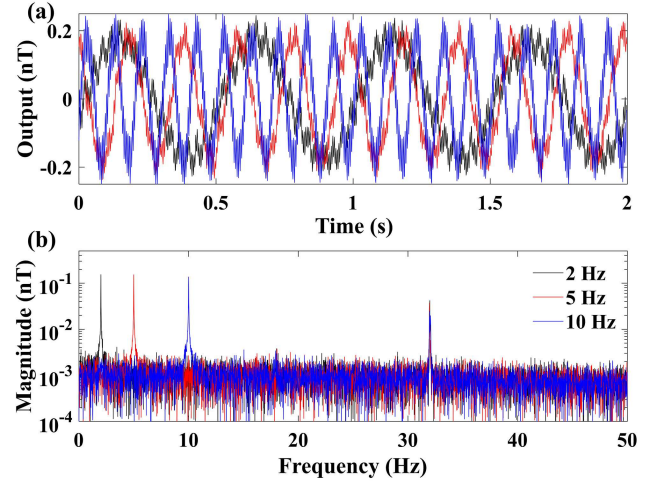}
	\caption{\label{sfig4} 
		(a) Output of the magnetometer (2 s cut) with 150 pT modulated test fields of 2 Hz, 5 Hz, and 10 Hz. The modulation frequency is 182 Hz.
		(b) is the Fourier transform of the acquired signals. The 32 Hz noise spike originates from the 150 Hz harmonics signal with the 182 Hz demodulation. The magnitude is plotted in logarithm. The noise level is as same as the noise at 182 Hz in Fig.5 of about 2 pT with the frequency resolution of 13.7 mHz.}
\end{figure}
Since the resonant MW frequencies of the NV spin change with magnetic field following $\Delta f=\gamma \Delta B$, the diamond magnetometer response to the external magnetic field can be acquired by sweeping the applied MW frequency, as shown in S-Fig. \ref{sfig1}. From the measured slope $k_c$, the transfer function of the magnetometer can be decided as $\Delta B=\Delta U/\gamma k_c$, where $\Delta U$ is the output voltage of the magnetometer. The intrinsic noise of the magnetometer can then be determined by the Fourier transform of the signal trace. In Fig. 5(a), the calculated spectrum is compared to the output spectrum of the OPM. In the comparison measurement, the diamond and CPC lens structure is replaced by the OPM sensor head. The other parts of the setup, i.e. laser, MW, photodetectors, keep working during the measurement. By this means, we make sure the OPM measures almost the same magnetic field noise as the diamond magnetometer. The only difference could be the coil noise. The coil noise includes the Johnson noise and the current shot noise of the batteries. The internal resistance of the battery is 16 m$\unit\Omega$ when it is fully charged. The resistance of each coil is 49.5 $\unit \Omega$. Thus, the Johnson noise of the coil is estimated as
\begin{equation}
	U_J=\sqrt{4kT(R_{coil}+R_{battery})\Delta\nu}
	\label{s20},
\end{equation}
where $k$ is the Boltzmann constant, $T$ is the temperature taken as 20 $^\circ$C, and $\Delta\nu=1$ Hz is the measured bandwidth. The current shot noise is calculated according to
\begin{equation}
	i_{shot}=\sqrt{2qI\Delta\nu}
	\label{s21},
\end{equation}
where $I$ is the current in the coil. Since the coil pair generates roughly $B_{bias} = 10^{-3}$ T field with $U$ = 12 V power supply, the magnetic field noise generated by the coil pair can be estimated as $B_{n,coil}=\frac{\sqrt{2}B_{bias}}{2U}\sqrt{U_J^2+(i_{shot} R_{coil} )^2}\approx0.8$ pT with 1 Hz bandwidth, and the noise level is 0.01 pT with measurement time of 73 s. Thus, the coil noise can be neglected with the fully charged batteries, and the noise spectral measured by the OPM and the diamond magnetometer are comparable to each other.

Another method to determine the intrinsic noise of the magnetometer is to apply a known oscillating field to the magnetometer and compare the detected signal magnitude at the frequency in the spectrum. One problem of the calibration method is that the known field signal spike could be underestimated because the magnetometer output includes both signal and noise. The underestimation of the signal magnitude leads to the underestimation of the detected noise floor. Thus, calibration measurements with known fields are used as the optional proof of the magnetometer characteristics.

The calibration field has a magnitude of 150 pT, and the frequencies are at 2 Hz, 5 Hz, and 10 Hz. In order to avoid the higher noise level and all the spikes at low frequencies according to the measured noise spectrum in Fig 5, the signals are mixed with a 182 Hz carrier and are generated by an arbitrary waveform generator. The modulated signal is applied to the coils. With the lock-in demodulation at 182 Hz we can directly see the low-frequency signal detected by the diamond magnetometer, as shown in S-Fig. 4(a). In order to suppress the harmonics, the cut-off frequency of the lock-in filter is set at 49 Hz, while the 32 Hz signal can still be seen due to the demodulation of 150 Hz noise spike at 182 Hz. The photodetector output is first translated to magnetic field according to the slope determined by sweeping the MW frequency. The resolved magnetic field magnitudes are 153.3 pT, 152.1 pT, and 136.8 pT at the three frequencies, respectively. The detected 10 Hz signal is significantly lower than 150 pT. The reason could be that the applied mixed signal has 172 Hz and 192 Hz components, and the 192 Hz signal is close to the 200 Hz bandwidth of the magnetometer, which is set by the cut-off frequency of the lock-in filter for demodulating the MW modulation. The magnetometer sensitivity of the three measurements can be calculated with the spectrum according to reference \cite{RN3}, as 14.5 pT/Hz$^{1/2}$, 16.7 pT/Hz$^{1/2}$, and 17.3 pT/Hz$^{1/2}$. The sensitivity results to a noise level of about 2 pT with the 73 s measurement, which is also the noise level at 182 Hz in the spectrum of Fig. 5.

\subsection{Modulated MW bandwidth}
\begin{figure}[htb]
	\includegraphics{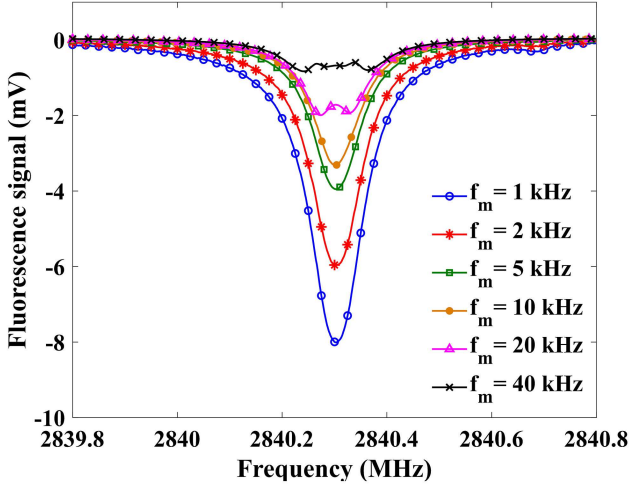}
	\caption{\label{sfig5} AM signal driving ODMR spectrum detected by LIA with the different amplitude modulation frequencies}
\end{figure}

One of the issues using the diamond with narrow linewidth is that the modulated MW could have a bandwidth similar to the original ODMR linewidth, and the modulated MW would reduce the ODMR contrast and broaden the linewidth. To look into the effect of applying modulated MW with a broaden bandwidth, we use an amplitude modulated (AM) MW signal with only two single-tuned sideband signals at $f_0+f_m$ and $f_0-f_m$. S-Figure \ref{sfig5} plots the ODMR spectrum driven by AM MW signals modulated at different frequencies. The contrast reduces with the increasing $f_m$ mainly because of the bandwidth of the magnetometer. There is an apparent splitting of the spectrum which further deteriorates contrast when the MW bandwidth $2f_m$ approaching the HWHM of the ODMR spectrum (from $f_m$ = 20 kHz). As for FM and PM signal, the bandwidth is defined by the Carson bandwidth:
\begin{eqnarray}
	&BW_{FM}=2(f_d+f_m)
	\label{s22},&\\
	&BW_{PM}=2(\phi_d+1)f_m
	\label{s23}.&
\end{eqnarray}

In this work, the applied modulation frequency $f_m$ = 9 kHz, and $\phi_d$ = 2. The Carson bandwidth $BW_{FM}$ = 54 kHz near the FWHM linewidth. The components amplitude is decomposed by the Bessel function first kind $J_n(\beta)$, which are 0.224, 0.577, 0.353, 0.129 for n=0 to 3. Thus, most of the MW energy (n=0 to 2) within the NV linewidth makes the parameters reasonable in the measurements.

\subsection{Enhancement with the flux guide}
\begin{figure*}[htb]
	\includegraphics[scale=0.9]{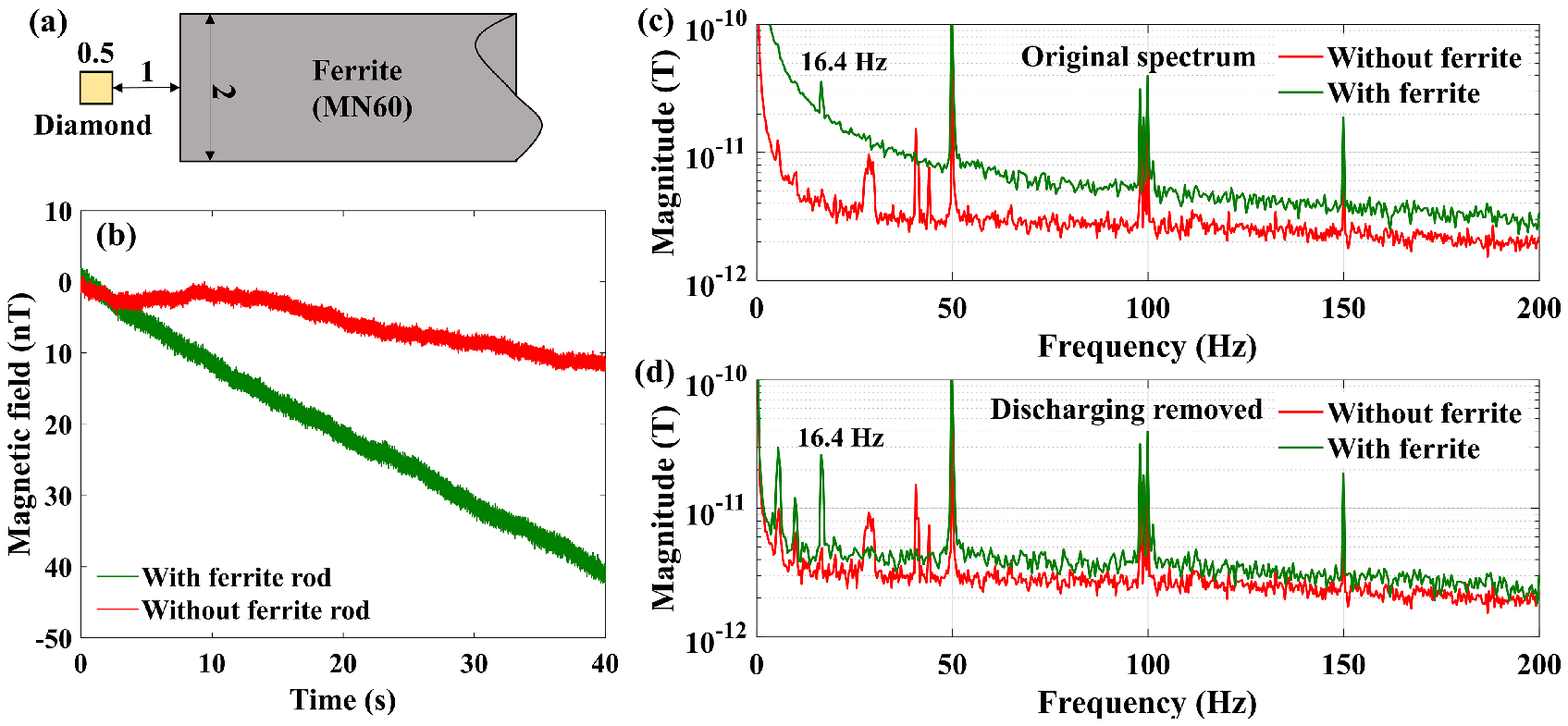}
	\caption{\label{sfig6} 
		(a) Key dimensions of the ferrite-diamond geometry used in experiment and flux simulation. The unit is in mm.
		(b) The magnetic field changes detected by diamond magnetometer with and without the FG. The ramping signals are induced by discharging of the batteries, and are used for calibration by linear fitting the discharging rate. 
		(c) The magnetic field noise spectra analyzed from the raw output of the magnetometer. 
		(d) The magnetic field noise spectra analyzed from the signal with the discharging field signals removed. The low-frequency noise spikes appear after the filtering.}
\end{figure*}
Schematic of the ferrite-diamond geometry is depicted in S-Fig \ref{sfig6}(a).  The magnetic field generated by the coil is concentrated and guided to the diamond through the FG. The main dynamic of the measured signal is the field change generated by the voltage dynamic of the batteries. In applications, getting closer to the magnetic field source is the alternation for having higher sensitivity. Thus, applying the flux guide to approach the magnetic field source is equivalent to an improvement of sensitivity.The raw outputs of the diamond magnetometer are plotted as S-FIG. \ref{sfig6}(b) shows. The field keeps reducing because the batteries are discharging. Detected field reduction rates are calculated by linear fitting the measured traces. The detected field reduction rate with FG is 6.3 times faster than the result measured without FG, which means the diamond magnetometer can detect 6.3 times weaker magnetic source from the other end of the FG than measurement without it. It is equivalent to having the scalar factor improved by 6.3 times. A geometry-based simulation is made by using Finite Element Method Magnetics (FEMM 4.2) software. The length of the ferrite rod and the radius of the coils used in the model is 16 cm, approximating the experimental parameters. The flux enhancement at the diamond center is calculated to be roughly 14 times. The reason for the enhancement difference to the experimental result could be errors in the primary parameters shown in S-Fig. \ref{sfig6}(a) which can affect the flux concentration a lot. In order to quantify the collection of extra noises by the flux guide, the discharging field is removed before we do Fourier transform to the signal traces.

S-Figure \ref{sfig6}(c) is the original noise floor spectra, and (d) is the spectra after removing the discharging signals from the outputs. The low-frequency noise spikes can be seen after removing the batteries discharging signal, and they are generated by the cooling fan motors of instruments in the laboratory. The low-frequency spikes are hard to be seen by the magnetometer in the shields. However, after applying the FG, of which one end is close to a window of the shields, the noise spikes are guided by the FG and detected by the magnetometer.  It can also be seen that more noises are collected by the ferrite rod, and the noise floor is lifted to 3 – 4 pT at the location of the diamond. Taking the signal gain induced by the ferrite rod into consideration, the SNR of the probe is improved, and the minimum detectable field is estimated as 0.3 - 0.7 pT according to the spectrum in Fig 5(b). 

\begin{figure*}[htb]
	\includegraphics[scale=1]{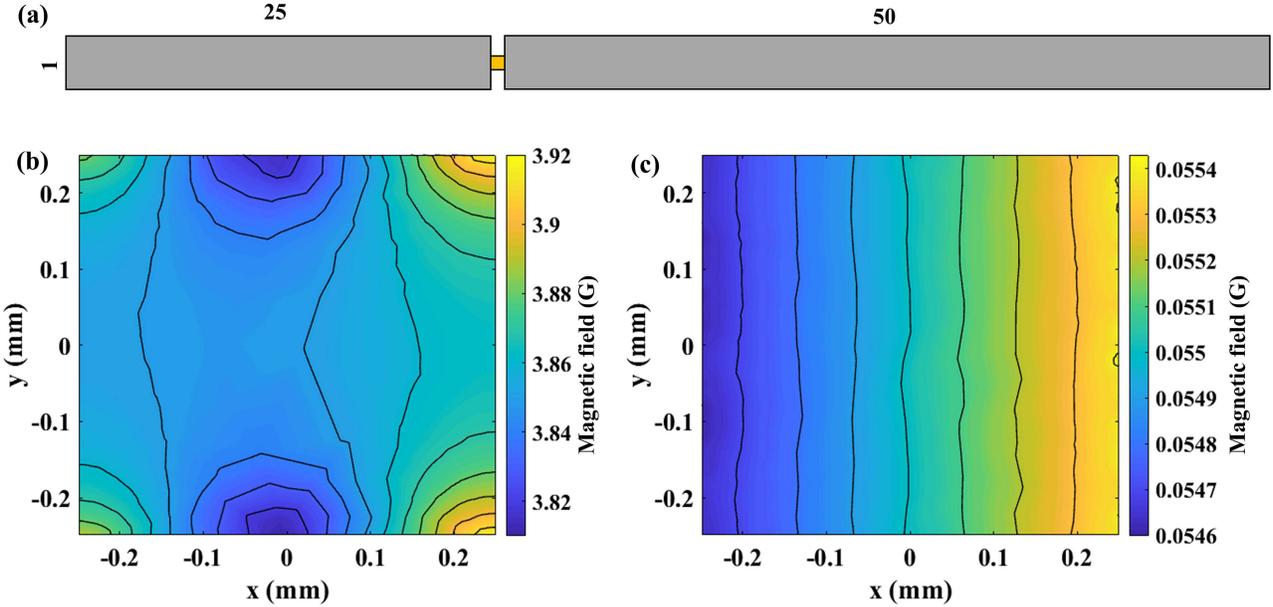}
	\caption{\label{sfig7} 
		(a) Schematic of a FG design, unit in mm. The shorter ferrite used a better field homogeneity around the diamond. A magnet dipole is placed 75 mm away from the diamond and 25 mm away from the ferrite rod end in the simulation. 
		(b) is the calculated field strength distribution with the FG.
		(c) is the field strength distribution calculated without the FG.}
\end{figure*}

The flux guide is not optimal in this experiment. For example, the gap between the ferrite and the diamond is as large as 1 mm, and it makes the flux gain much lower than using a smaller gap. A flux return ferrite could give a higher gain for flux concentration and better field homogeneity. We simulate a flux guide schematic as S-FIG. \ref{sfig7} draws, and it is a planar simulation. A magnet dipole is placed 75 mm away from the diamond and the field at location of the diamond without FG is 0.055 G. The difference of the field distribution within the diamond is 0.4\% according to the calculation. When the flux guide is applied, the flux is concentrated and guided, and the magnetometer detects a field of 3.86 G, with the field difference also around 0.4\%. Thus, the field is amplified by 70 times in the planar simulation, while the spatial resolution is kept around 1 mm according to the distal diameter of the FG.

\end{document}